\def\tzero{$t^\mathrm{FRB}_{0}$}

\documentclass[preprint]{aastex61}
\usepackage{lineno}

\newcommand\aastex{AAS\TeX}

\received{Nov 21, 2020}
\revised{-}
\accepted{\today}
\submitjournal{ApJ}

\shorttitle{\aastex\ FRB 131104}
\shortauthors{Sakamoto et al.}

 \usepackage{color}

\begin{document}
\title{FRB131104 Swift BAT data revisited: 
No evidence of a gamma-ray counterpart}

\author{T.~Sakamoto}
\affil{College of Science and Engineering, 
Department of Physics and Mathematics, 
Aoyama Gakuin University, 
5-10-1 Fuchinobe, Chuo-ku, 
Sagamihara-shi Kanagawa 252-5258, 
Japan}

\author{E.~Troja}
\affil{NASA Goddard Space Flight Center, Greenbelt, MD 20771, USA}
\affil{Department of Astronomy, University of Maryland
College Park, MD 20742, USA}

\author{A.~Lien}
\affil{Department of Physics, University of Maryland, Baltimore County, 1000 Hilltop Circle, Baltimore, MD 21250, USA}
\affil{Center for Research and Exploration in Space Science and Technology (CRESST) and 
NASA Goddard Space Flight Center, Greenbelt, MD 20771, USA}

\author{B.~Zhang}
\affil{Department of Physics and Astronomy, University of Nevada Las Vegas, NV 89154, USA}

\author{S.~B.~Cenko}
\affil{NASA Goddard Space Flight Center, Greenbelt, MD 20771, USA}

\author{V.~Cunningham}
\affil{Department of Astronomy, University of Maryland College Park, MD 20742, USA}

\author{E.~Berger}
\affil{Harvard-Smithsonian Center for Astrophysics, 60 Garden Street, Cambridge, 
MA 02139, USA}

%
%
%
%



\begin{abstract}

We present a re-analysis of the \textit{Swift} Burst Alert Telescope (BAT) data around the radio detection of FRB~131104. Possible evidence of a gamma-ray counterpart was presented by \citet{delaunay2016}.  However, based on our analysis using all the available BAT data, no significant emission is found in either the temporal or the image domain.  We place a 5~$\sigma$ fluence upper limit  of 3.3 $\times$ 10$^{-6}$ erg cm$^{-2}$ and 2.7 $\times$ 
10$^{-6}$ erg cm$^{-2}$
($15-150$ keV) with an integration time of 300 s assuming a simple power-law spectrum with photon index 
of $-1.2$ and $-2.0$, respectively.   
Our result does not support the association of this FRB with a high-energy counterpart, in agreement with growing observational evidence that most FRBs are not associated to catastrophic events such as gamma-ray bursts. 

\end{abstract}

\keywords{radio transient sources, radio bursts, gamma-ray bursts, X-ray transient sources}



\section{Introduction} \label{sec:intro}

Fast Radio Bursts (FRBs; \citealt{lorimer2007, thornton2013}) are bright ($\approx$Jy) and short-lived ($\lesssim$10~ms) pulses of coherent, 
low-frequency ($\approx$0.4--8~GHz) radio emission. 
With an estimated all-sky rate of $\approx$10$^{3}$ - 10$^{4}$~events per day \citep{thornton2013,champion2016,connor2016},
FRBs dominate the transient radio sky. Yet their true nature remains a mystery. 
Mounting observational evidences -- such as their high Galactic latitudes, excessive dispersion and rotation measures
 \citep{lorimer2007,thornton2013,masui2015,petroff2016}, and the recent identification of their host galaxies \citep[e.g.][]{chatterjee2017,Bhandari2020} --
suggest that FRBs originate from cosmological distances. They have a power law luminosity function spanning $\sim 8$ orders of magnitude \citep{luo2018,luo2020a,lu2020a,lu2020b},
and an isotropic equivalent energy in the range $\sim (10^{35}-10^{43})$ erg \citep{zhang2020}. 

The fact that at least some FRB sources repeat \cite{spitler2016,scholtz2016,chime2019a,chime2019b,luo2020b} and that the FRB event rate density exceeds those of catastrophic explosions such as supernovae \citep{ravi2019,luo2020a} suggest that the majority of FRBs should arise from non-catastrophic events. Likely scenarios include bursting magnetar \citep{popov2013,lyubarsky2014,katz2016,lu2020b} or radio pulsar giant pulses \citep{cordes2016,connor2016b}, or some models invoking interactions with a neutron star magnetosphere \citep{zhang2017,zhang2020}. 
Nevertheless, the possibility that a small fraction of FRBs may be related to catastrophic events is not ruled out.

Among catastrophic models, several scenarios predict an association between an FRB and a gamma-ray burst (GRB). First, if the collapse of a supramassive neutron star can make an FRB \citep{falcke2014},
\citet{zhang2014} showed that a GRB could be followed by an FRB, occurring at the time when the internal X-ray plateau abruptly ends \citep{Troja2007}. 
Second, several authors have suggested that compact binary mergers may produce an FRB in association with a short GRB \citep[e.g.][]{wang2016,zhang2016}. Finally, even for non-catastrophic events, a foreground neutron star magnetosphere being suddenly reshaped by the blastwave of a GRB can also make an FRB associated with a GRB \citep{zhang2017}. 
A robust GRB-FRB association is essential to constrain these models. 

Most searches for coincident GRB-FRB associations, unsurprisingly, gave negative results 
\citep[e.g.][]{bannister2012,palaniswamy2014,sun2018,cunningham19,martone2019,anumarlapudi2020,guidorzi2020,palliyaguru2020,rowlinson2020}. 
However, one potential association has been claimed. 
\citet{delaunay2016} (hereafter D16)  presented the discovery of a long-lived ($\approx$380~s) hard X-ray transient, Swift J0644.5-5111, spatially coincident with FRB~131104 \citep{ravi2015}. 
The onset of the high-energy emission appears delayed by $\approx$300~s with respect to the FRB, and due
 to its low significance ($\lesssim$4$\,\sigma$)  it  was not found by the standard on-board trigger algorithm, but during a targeted archival search for FRB counterparts. This search included two more known FRBs for which no gamma-ray counterpart was detected in the \textit{Swift} data.
A possible origin for Swift J0644.5-5111 and its association to FRB131104 was discussed by \cite{murase2017}. \cite{shannon2017} pointed out that the lack of a radio afterglow disfavors a standard GRB fireball scenario, whereas \cite{gao2017} found that the radio upper limits can be still consistent with the existence of a GRB for a reasonable range of parameters. \cite{zhang2017} suggested that the event could be a result of interaction between a GRB blastwave and the magnetosphere of a foreground neutron star (the ``cosmic comb'' model). 

The association between a long-lasting high-energy transient and an FRB, if confirmed, would greatly constrain FRB source models, at least for this event. 
For instance, the reported long duration disfavors those scenarios with a short-lived central engine, such as compact object mergers. 
Most importantly,  if placed at the same estimated redshift $z$\,$\sim$0.5 of FRB~131104, 
the observed fluence $S_{\gamma} \approx 4 \times 10^{-6}$ erg cm$^{-2}$  implies a total energy release
$E_{\gamma,iso} \approx 5 \times 10^{51}$ erg, similar to GRBs.
This would challenge most non-catastrophic progenitor models and suggest that there could be a category of catastrophic FRBs.

However, it should not be overlooked that the claimed detection of D16 is suspiciously located at the edge of the Burst Alert Telescope (BAT)'s field of view.
This region, characterized by a low ($\approx$5\%) partial coding,  is known to be affected by large noise fluctuations and complex systematic effects, and is generally excluded in standard analyses. 
For this reason, we performed a re-analysis of the \textit{Swift}/BAT data presented by D06 
and assessed the significance of the putative source Swift J0644.5-5111 with different analysis techniques.

Throughout the paper we use as reference time \tzero\ the radio detection time of FRB~131104, which is 18:03:59 UT on November 4th, 2013 \citep{ravi2015}.

\section{BAT Data Analysis}

At the FRB trigger time \textit{Swift} was pointing toward the location of GRB~130925A \citep{piro2014},
at R.A., Dec. = 41.105$^{\circ}$, $-26.135^{\circ}$ (J2000).
\textit{Swift} slewed to this pre-planned target starting from \tzero $-$171 s, settled at \tzero $-$19 s,
and kept observing it until 18:17:57 UT ($\sim$\tzero+14 min).
No BAT on-board trigger occurred at or around \tzero.  For this reason, the typical burst data products, such as the event data
with full time and spectral resolution, are not available. 
However, the BAT produces several other data products all the time, regardless of whether there is a trigger or not. 
For our search, we used these continuous BAT data, such as the raw light curves data (Sec.~\ref{raw}), 
the survey data (Sec.~\ref{survey}) and the scaled map data (Sec.~\ref{scaled}).  

\subsection{Raw light curves data}\label{raw}

\begin{figure}
\centerline{
\includegraphics[scale=0.35,angle=0]{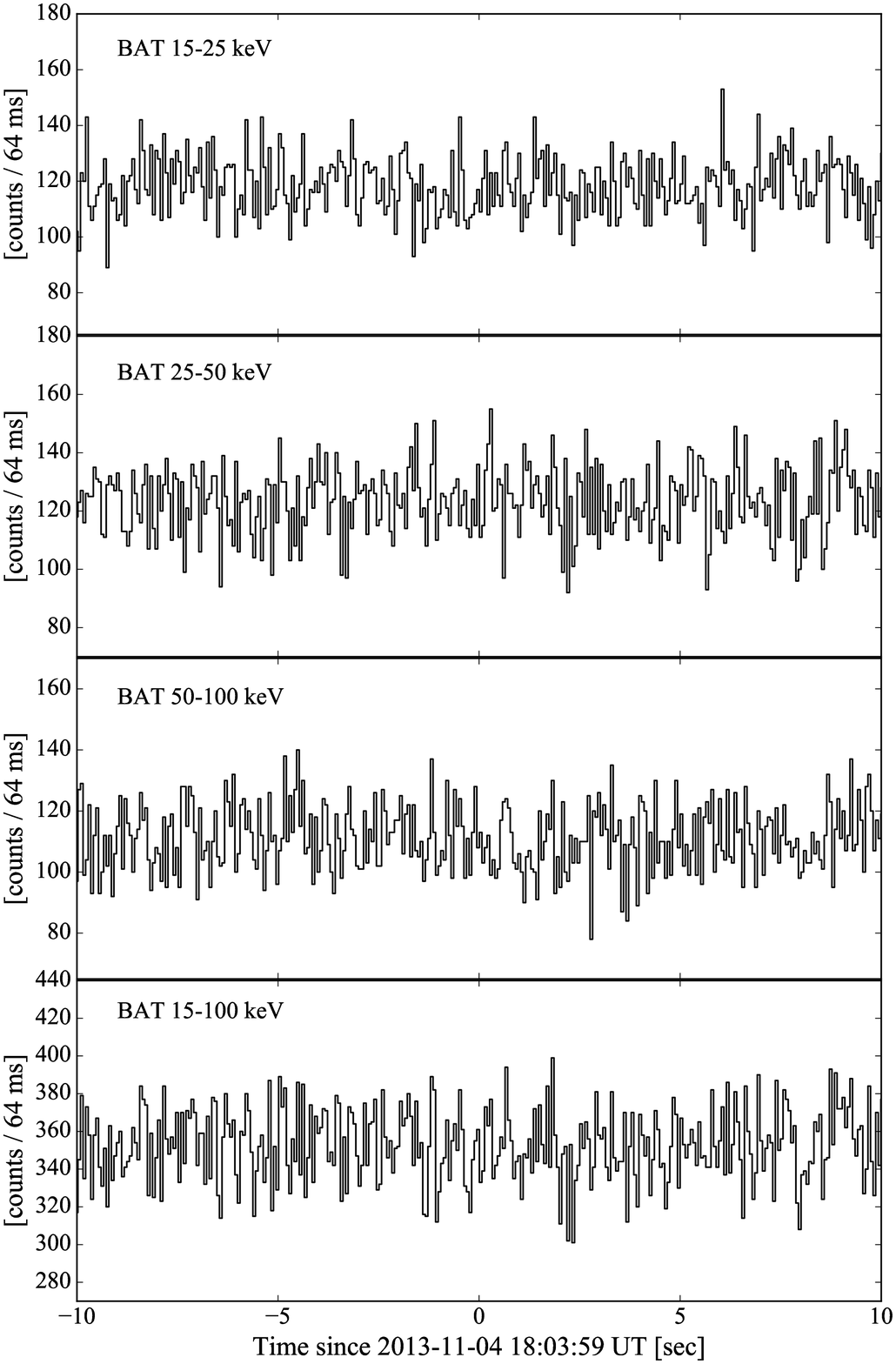}
\includegraphics[scale=0.35,angle=0]{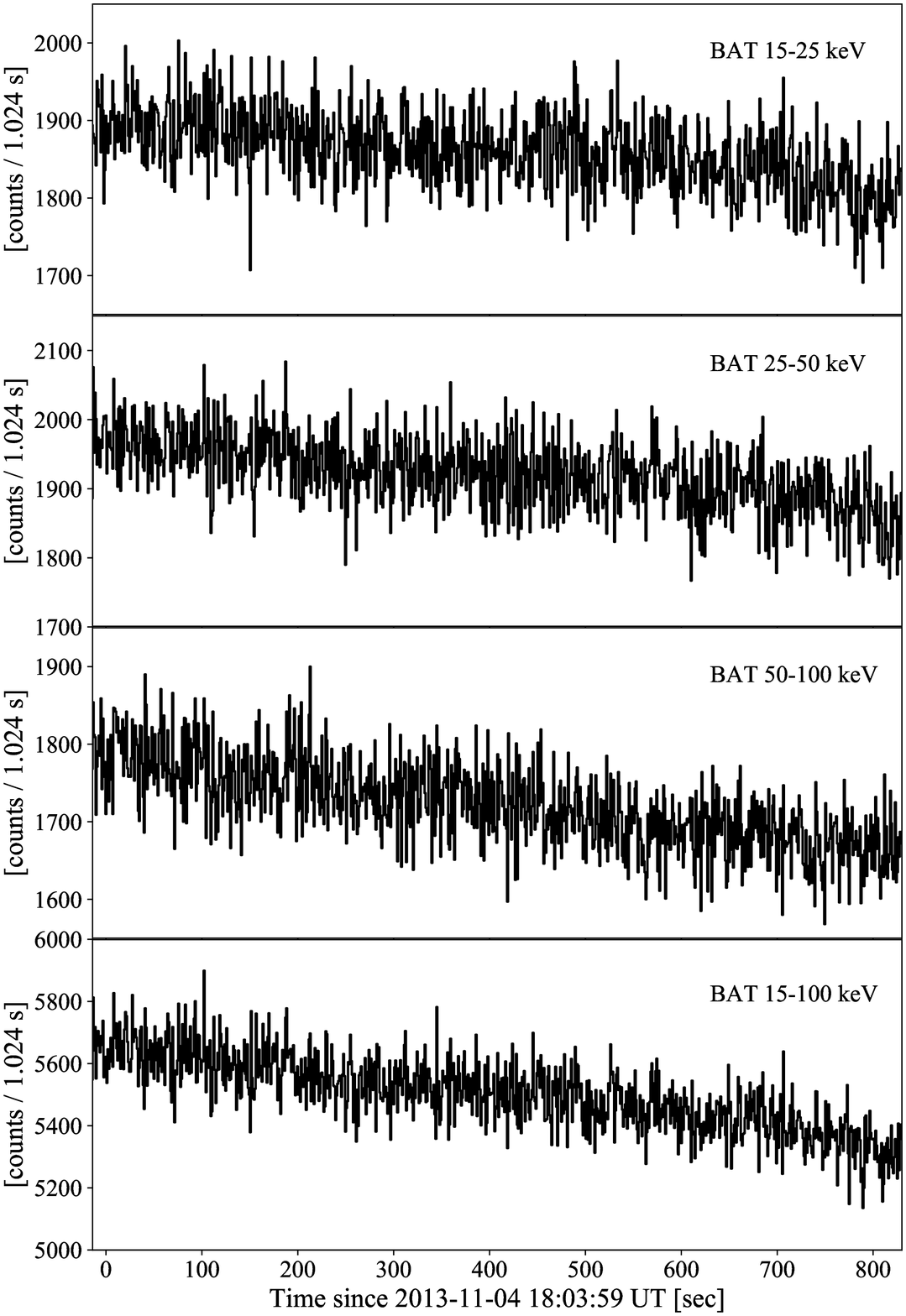}}
\caption{The BAT 64 ms (left) and 1.024 s (right) raw light curves in the 15-25 keV, the 25-50 keV, the 50-100 keV 
and the 15-100 keV bands at \tzero $\pm$ 10 s (left) and from \tzero  $-19$ s to \tzero +14 min (right).}
\label{fig:BAT_64ms_lc}
\end{figure}

Figure \ref{fig:BAT_64ms_lc} shows the  64 ms and 1.024 s raw (i.e. non-background subtracted)
light curves around \tzero\ in the four energy bands (15-25 keV, 25-50 keV, 50-100 keV and 15-100 keV).  
Each light curve is accumulated over the full array of BAT detectors and has no positional information.
Since the distribution of the count rate at the 100-350 keV band shows a systematically larger standard 
deviation than that of the rest of the light curves, we removed the data of the 100-350 keV  band from the search.  
We searched for excess emission above the background level over the time interval from \tzero $-$19~s to 
\tzero $+$14~min when FRB 131104 was in the field of view of BAT. 
The signal-to-noise ratio (SNR) was calculated by scanning the light curve bins using the foreground count in one time 
bin and the background count in ten time bins prior to the foreground bin.   
The searched timescales are 64~ms, 128~ms, 512~ms, 1.024~s, 2.048~s, 5.120~s, and 10.24~s.  
The detection thresholds are set as 3 $\sigma$ for 64~ms and 128~ms, 2 $\sigma$ for 512~ms, 1.024~s, and 2.048~s, 
1 $\sigma$ for 5.120~s and 0.05 $\sigma$ for 10.24~s timescales. 
Table \ref{tbl:snr_raw_lc_search} summarized the maximum SNR and also calculated false-positive 
detection applying the number of trial in four different energy bands and seven different time scales.  
Given the high chance of a false positive detection, we concluded that no 
significant excess was found between 64~ms and 10~s time scales in the BAT raw light curve data.

\subsection{Survey data}\label{survey}
BAT survey data are continuously collected during pointed observations. They have typical integration times of 5 minutes
and are stored in an 80 energy channel histogram for each detector.
These data are used, for example, in the BAT 
all-sky hard X-ray survey \citep{markwardt2007,tueller2010,baumgartner2013} and also for searching 
a weak GRB emission prior to and post to the GRB trigger time \citep{lien2016}.  
They are particularly useful to investigate a weak and long-lasting transient as they allow for a good characterization of the sky background over the same temporal window. 

\begin{figure}
\centerline{
\includegraphics[scale=0.8,angle=0]{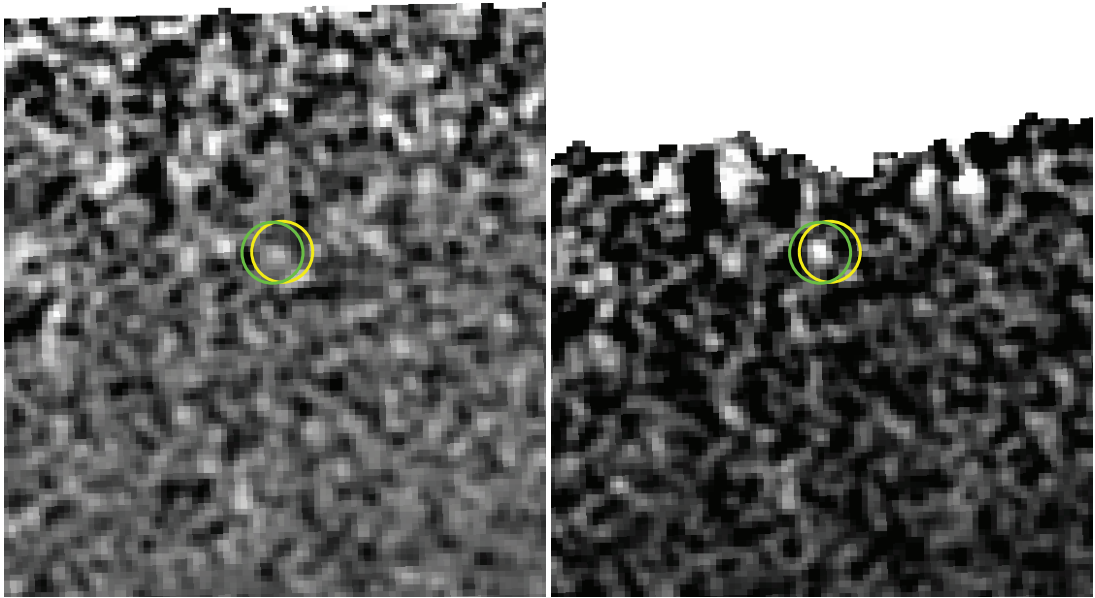}}
\caption{Left: The BAT sky image in the 15-150 keV band generated from the survey data 
specifying the `detection mask' in the standard analysis.  Right: The BAT sky image in 
the 15-150 keV band generated from the survey data specifying the `flux mask' in the analysis.  
The FRB position is shown by the yellow circles, whereas the position of the candidate counterpart is shown by the green circles.   
The integration time interval of the images is between \tzero$-7$ s and \tzero$+273$ s (300 s integration time).} 
\label{fig:BAT_survey_img}
\end{figure}

In our case, survey data are available starting from \tzero$-$7 s.  
We processed them using the standard {\tt batsurvey} script.  As described in the \textit{Swift} BAT 
calibration document, SWIFT-BAT-CALDB-CENTROID-v2\footnote{http://heasarc.gsfc.nasa.gov/docs/heasarc/caldb/swift/docs/bat/SWIFT-BAT-CALDB-CENTROID-v2.pdf}, three types of the aperture masks are included in the BAT calibration database (CALDB).  
They are the full aperture mask (detection mask), the reduced aperture mask (flux mask) 
and the difference between the full and the reduced aperture mask (edge mask).  
Although the `detection mask'  is the actual BAT aperture mask, 
it is usually recommended to use the `flux mask', which excludes the edges of the field of view. 
This is because fluxes for far off-axis sources are often not accurate and should not be considered in standard analyses.
However, in our case, the FRB position is at the edge of the field of view and the `detection mask' should be used to accurately measure the significance of detected sources.  
The source partial coding fraction using the `detection mask' is only 5.7\%.   
The left panel of figure \ref{fig:BAT_survey_img} shows the BAT sky image in the 15-150 keV band between \tzero$-7$ s and \tzero$+273$ s (total integration time of 300s). 
Furthermore, the {\it Swift} BAT team is using the `detection mask' for GRBs located at the partial coding 
fraction less than 10\% in the published three GRB catalog papers \citep{bat_grb1_cat,bat_grb2_cat,lien2016}.  Our work also uses the 'detection mask' which is the recommended analysis procedure for a source located at a low partial coding fraction. 

\begin{figure}
\centerline{
\includegraphics[scale=0.8]{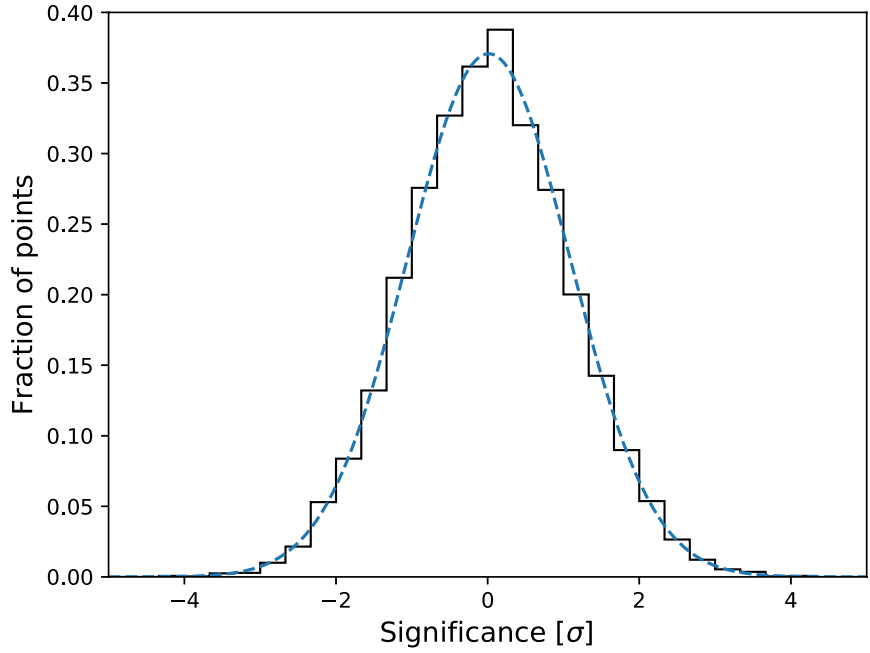}}
\caption{Histogram of pixels significance for random positions across the BAT sky image in the 15-150 keV band.  The dotted line is the best fit Gaussian function.}
\label{fig:BAT_15_150kev_sigifmap}
\end{figure}

The source significance at the radio\footnote{The radio position used in the analysis is (R.A., Dec.) = (101.040, $-51.271$) (J2000).} 
and the D16 positions is 1.7 $\sigma$ and 2.4 $\sigma$, respectively.  
To quantify the significance directly from the data, we randomly ran the source detection software, {\tt batcelldetect}, 
throughout the image and extracted the SNR at random positions.  We extracted 8379 random points in the BAT sky image, and the distribution of their significance is shown in
Figure \ref{fig:BAT_15_150kev_sigifmap}.  Both 1.7 $\sigma$ and 2.4 $\sigma$ values are well within the noise distribution of the image, 
and the probability to measure a comparable SNR at a random position is 6.1\% and 1.4\%, respectively.  Therefore, we conclude that the signal is not statistically significant to claim a detection in the BAT survey data.

The fluence upper limit is estimated as follows.  
From the noise sky map, we derive an average value of 3.2 $\times$ 10$^{-2}$ counts s$^{-1}$
at the radio FRB position, which corresponds to a
5 $\sigma$ detection threshold of $\approx$0.16 counts s$^{-1}$. 
Using the task {\tt batdrmgen}, 
we then generate an energy response file at the 
incident angle of the FRB position ($\approx$ 51$^{\circ}$) and input it into {\tt xspec}. 
The rate-to-flux conversion factor is derived
assuming a power-law spectral shape with photon index of $1.16$ (the photon index of the claimed D16 source) and $2.0$ (Crab-like spectrum).
 The resulting 5 $\sigma$ fluence upper limits in the 15-150 keV with an integration time of 300 s 
are 3.3 $\times$ 10$^{-6}$ erg cm$^{-2}$ and 2.7 $\times$ 10$^{-6}$ erg cm$^{-2}$, respectively.  The fluence reported by D16 is $\approx$4.0 $\times$ 10$^{-6}$ 
erg cm$^{-2}$ which is at a comparable level of our derived upper limit.  

The results of D16 can be reproduced only by processing the survey data using the `flux mask'.  The right panel 
of Figure \ref{fig:BAT_survey_img} shows the corresponding sky image.  In this image the significance of the candidate counterpart is 4.3 $\sigma$, close to what reported in D16.  Therefore, it is clear that the main
difference between D16 and our analysis is in the selection of the aperture mask, and that the significance of the high energy source sensitively
on this particular choice, dropping from 4.3 $\sigma$ (with the `flux mask') to 2.4 $\sigma$ (with the `detection mask').  
Based on this fact, we conclude that the statistical significance adopted by D16 overestimates the true SNR of this source, and find no robust evidence to claim a detection of a high-energy counterpart at the FRB position.


\subsection{Scaled map data} \label{scaled}

During regular {\it Swift} operations, the onboard processor stores detector plane count maps every 64 s
in order to enable the discovery of long duration transients (image trigger). The maps are collected in the 15-50 keV energy band, and then scaled relative to a full scale value of 255. 
These scaled maps can then be convolved with the 
mask pattern to produce a BAT sky image.

For FRB131104, the scaled map data are available from $t^{\rm FRB}_{0}-2$ s to  $t^{\rm FRB}_{0}$+830 s
with integration time of 64 s, 320 s, and a full 832 s exposure.  The scaled map data are processed 
using the standard BAT tools.  The scaled map count map is transferred to the sky image 
using {\tt batfftimage} specifying the `detection mask' aperture.  The generated sky image is cleaned 
only for the flat fielding using {\tt batclean} (the balancing option is `flight').  Then, the 
count rate at both the radio and the D16 positions are extracted by {\tt batcelldetect}.  Finally, the 
extracted count rates are corrected to the on-axis count-rates based on the Crab data 
\citep{bat_gsc_trans_mon}.  

\begin{figure}
\centerline{
\includegraphics[scale=0.4,angle=0]{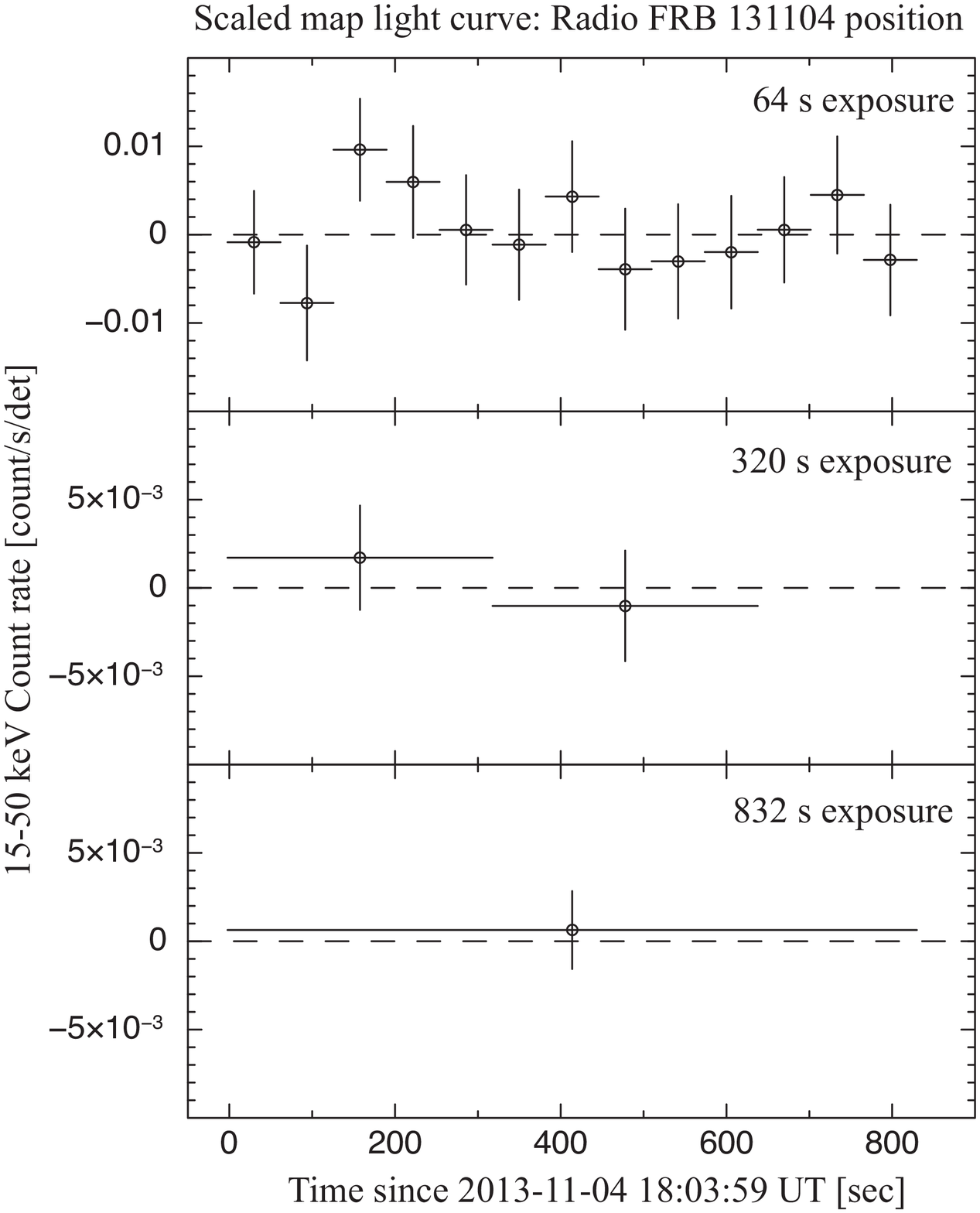}
\hspace{5mm}
\includegraphics[scale=0.4,angle=0]{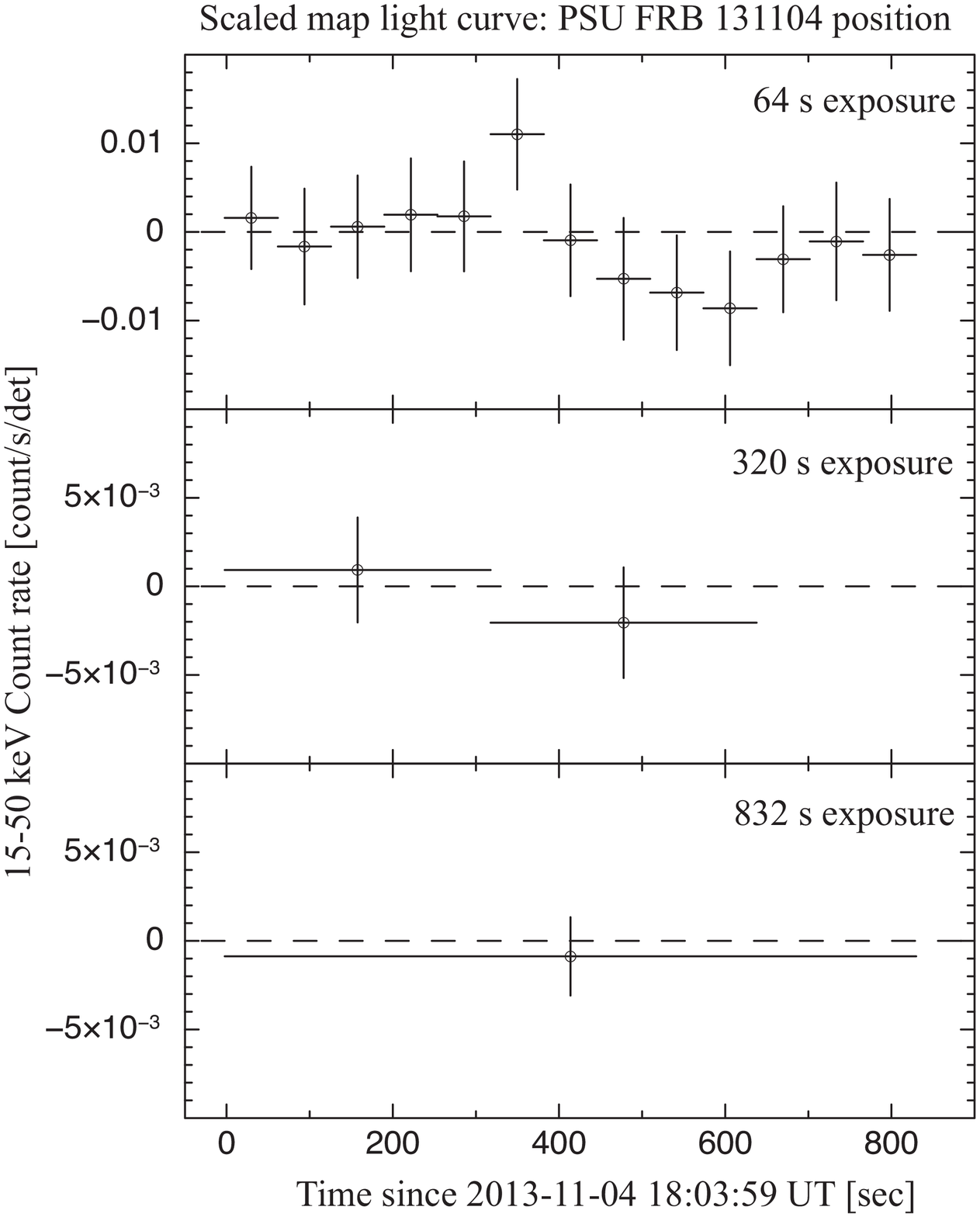}
}
\caption{The 15-50 keV light curves generated from scaled map data using the radio FRB position (left) 
and the D16 position (right).}
\label{fig:smap_lc}
\end{figure}

Figure \ref{fig:smap_lc} shows the light curves in the 15-50 keV 
band generated from the scaled map data using the radio and the D16 positions.  
No significant emission is evident, and the count rate   
is seen to fluctuate around zero (top panel of figure \ref{fig:smap_lc}).  During the time interval of the putative BAT transient ($t^{\rm FRB}_{0}$+318 s and $t^{\rm FRB}_{0}$+637 s), the 
source significance is 0.3 $\sigma$. 
For comparison, the light curve generated from the same scaled map data 
presented in D16 (Figure 1 of D16) shows a gradual trend (from high to low) in count rate, which is similar to the background trend seen in the right panel of Figure \ref{fig:BAT_64ms_lc}.

In order to improve the sensitivity of our search, 
we selected a background interval ($t^{\rm FRB}_{0}-2$ s and $t^{\rm FRB}_{0}$+318 s) and subtracted the corresponding image to the source image, 
extracted over the time interval $t^{\rm FRB}_{0}$+318 s and $t^{\rm FRB}_{0}$+637 s. 
In this subtracted image the source significance increases to 0.7~$\sigma$, and is therefore not sufficient for a detection.

\section{Conclusion and Discussion}
We investigated all the available BAT data around 
the onset of FRB~131104, from the raw light curves to the sky images.  No significant emission is seen in the search for both short timescales (64 ms and 1.6 s)  based on the light curve analysis and long timescales (64 s up to 5 min) based on the image analysis.  
Our null result does not confirm the detection of a hard X-ray counterpart for FRB 131104, claimed by D16 using the same BAT data.   

\begin{figure}
    \centering
    \includegraphics[scale=0.6]{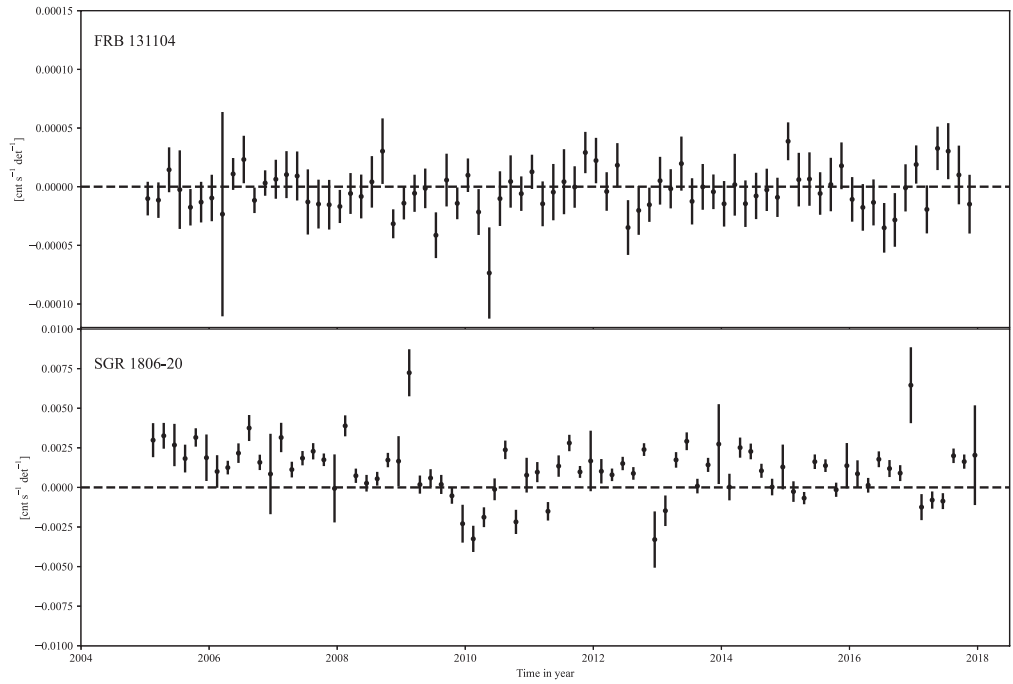}
    \caption{BAT survey light curves of FRB 131104 (top) and SGR 1806-20 (bottom) in the 14-195 keV band. The time bins are averaged over two months to increase the statistics.  No significant detection is evident at the position of FRB 131104 over the last 14 years.  Whereas the Galactic magnetar SGR 1806-20 shows a statistically significant detection.}
    \label{fig:bat_monthly_frb_sgr}
\end{figure}

As shown in Figure \ref{fig:bat_monthly_frb_sgr}, no significant detection from FRB 131104 is seen by BAT over the last 14 years of its operation.  On the other hand, the hard X-ray emission from the Galactic magnetar, SGR 1806-20, is constantly detected by the monthly averaged data of BAT.  This demonstrates that long lasting emission from an active magnetar can be seen by BAT and recovered by our analysis.  However, such an emission is not evident from FRB 131104.

Recently the FRB-like millisecond radio bursts and the burst in the hard X-ray band were simultaneously detected from the known Galactic magnetar SGR 1935+2154 \citep{mereghetti2020}.  This discovery 
provides strong support for a magnetar as the origin of at least some fraction of FRBs.  Although it is not possible to conduct a detailed search to investigate as a magnetar origin for FRB~131104 because of the limitation of the available data, a recent new capability of downlinking the BAT event data based on an external trigger by a rapid uplink commanding \citep{tohuvavohu2020} will enhance to identify such a possibility by the BAT data.

Our result suggests that so far there is no evidence of FRBs associated with catastrophic events such as GRBs. This is consistent with the consensus in the community that the majority of FRBs, if not all, are related to non-catastrophic, repeating sources \citep[e.g.][]{nicholl17,ravi2019,luo2020a}. The existence of catastrophic events remains possible, but the fraction should be quite small \citep{zhang2020}. Continued searches of FRBs in association with catastrophic events (GRBs and compact binary coalescence gravitational wave signals) will progressively tighten the upper limit of the fraction of catastrophic FRBs.

\acknowledgements
We would like to thank the anonymous referee for comments and suggestions that materially improved the paper.  
We also would like to thank D.~B. Fox and J.~J. DeLaunay for valuable comments and discussions.  This work is 
supported by MEXT KAKENHI Grant Numbers 17H06357 and 17H06362.

\begin{table}[t]
\centering
\caption{Summary of the raw light curve search.  The highest significance of a time bin and its time from the trigger time are 
shown in the upper raw.  The expected false-positive detection by the given number of trial is shown in the 
lower row. \label{tbl:snr_raw_lc_search}}
\begin{tabular}{c|c|c|c|c|c}
Searched time scale [s] & \multicolumn{4}{c|}{Maximum SNR \& False-positive detection} & \# of trial\\\cline{2-5}
 & 15-25 keV & 25-50 keV & 50-100 keV & 15-150 keV &\\\hline
0.064 & 3.8 $\sigma$ (153.8 s) & 4.2 $\sigma$ (383.2 s) & 4.3 $\sigma$ (234.8 s) & 4.0 $\sigma$ (70.3 s) & 13421\\
          & 0.97 & 0.18 & 0.11 & 0.43 &\\\hline
0.128 & 3.8 $\sigma$ (605.1 s) & 3.6 $\sigma$ (539.6 s) & 4.1 $\sigma$ (169.1 s) & 4.2 $\sigma$ (13.8 s) & 6710\\
          & 0.49 & 1.1 & 0.14 & 2.3 &\\\hline
0.512 & 3.1 $\sigma$ (89.0 s) & 2.9 $\sigma$ (280.0 s) & 3.2 $\sigma$ (637.4 s) & 3.4 $\sigma$ (346.1 s) & 1677\\
          & 1.6 & 3.1 & 1.2 & 0.56 &\\\hline
1.024 & 2.9 $\sigma$ (76.2 s) & 3.3 $\sigma$ (187.8 s) & 3.5 $\sigma$ (213.4 s) & 3.3 $\sigma$ (102.9 s) & 838\\
          & 1.6 & 0.41 & 0.19 & 0.4 &\\\hline
2.048 & 3.6 $\sigma$ (488.9 s) & 3.0 $\sigma$ (187.8 s) & 3.9 $\sigma$ (212.4 s) & 3.0 $\sigma$ (187.8 s) &  419\\
          & 0.067 & 0.57 & 0.020 & 0.57 &\\\hline
5.120 & 1.7 $\sigma$ (525.8 s) & 1.4 $\sigma$ (525.8 s) & 0.1 $\sigma$ (489.9 s) & 1.3 $\sigma$ (525.8 s) & 167\\
          & 7.4 & 13 & 76 & 16 &\\\hline
10.240 & 0.4 $\sigma$ (484.8 s) & $<$0.1 $\sigma$ & $<$0.1 $\sigma$ & $<$0.1 $\sigma$ & 83\\
            & 28 & $>$38 & $>$38 & $>$38 &\\\hline
\end{tabular}
\end{table}

%



\newpage

\newpage


\end{document}